\documentclass[12pt,preprint]{aastex}

\newcommand{\etal}{{\it et al. }}

\slugcomment{To appear in Astrophysical Journal} 
\slugcomment{}

\shorttitle{SN 2001gd}
\shortauthors{Stockdale, \etal}

\received{2003 March 14}
\begin{document}

\title{Radio Emission from SN~2001gd in NGC~5033} 
\slugcomment{To appear in Astrophysical Journal}

\author{Christopher J. Stockdale\altaffilmark{1}}
\affil{Naval Research Laboratory, Code 7213, Washington, DC 20375-5320; stockdale@nrl.navy.mil}

\author{Kurt W. Weiler}
\affil{Naval Research Laboratory, Code 7213, Washington, DC 20375-5320; Kurt.Weiler@nrl.navy.mil}

\author{Schuyler D. Van Dyk}
\affil{IPAC/Caltech, Mail Code 100-22, Pasadena, CA 91125; vandyk@ipac.caltech.edu}

\author{Marcos J. Montes}
\affil{Naval Research Laboratory, Code 7212, Washington, DC 20375-5320; Marcos.Montes@nrl.navy.mil}

\author{Nino Panagia\altaffilmark{2}}
\affil{Space Telescope Science Institute, 3700 San Martin Drive,
Baltimore, MD 21218; panagia@stsci.edu}

\author{Richard A. Sramek}
\affil{National Radio Astronomy Observatory, P.O.~Box 0, Socorro, NM 87801; dsramek@nrao.edu}

\author{M. A. Perez-Torres}
\affil{Instituto de Astrof\'{\i}sica de Andaluc\'{\i}a, CSIC, 
Apdo. Correos 3004, 18080 Granada, Spain; torres@iaa.es}

\and

\author{J. M. Marcaide}
\affil{Depto. de Astronomia Univ. Valencia, 46100 Burjassot, Spain; J.M.Marcaide@uv.es}

\altaffiltext{1}{National Research Council Postdoctoral Fellow.}
\altaffiltext{2}{On assignment from the Astrophysics Division, Space Science Department of ESA.}

\begin{abstract}

We present the results of monitoring the radio emission
 from the Type IIb supernova SN 2001gd 
between 2002 February 8 and 
2002 October 28.  Most of the data were obtained using the
Very Large Array at the five wavelengths of $\lambda \lambda$1.3 cm (22.4 GHz), 
2.0 cm (14.9 GHz), 3.6 cm (8.44 GHz),
6.2 cm (4.86 GHz), and 21 cm (1.4 GHz).  Observations were also made with 
Giant Meterwave Radio Telescope at $\lambda$21 cm (1.4 GHz).
The object was discovered optically
well after maximum light,
making any determination of the
early radio evolution difficult.  However, subsequent observations indicate that 
the radio emission has evolved regularly in both time and frequency and
is well described by the SN shock/circumstellar medium interaction model.  
\end{abstract}

\keywords{galaxies: individual (NGC~5033) -- radio continuum: stars  -- 
stars: mass-loss -- supernovae: general -- supernovae: individual (SN~2001gd)}

\section{Introduction}

Supernova (SN)~2001gd in NGC~5033 was discovered at magnitude, 
$16{\fm}5$ on 2001 November 24.820 UT by 
Nakano \etal (2001) at 
RA(J2000) = $13^h 13^m 23\fs89$, Dec(J2000) = $+36\arcdeg 38\arcmin 17\farcs7$,
with independent discoveries by other observers following quickly.   No source had been
 detected  at  that position (limiting mag. $\sim$ 18.5) on a previous 
image taken on 2001 May 8 and,
nothing is visible on the Palomar Sky Survey red and blue plates (Nakano \etal 2001).
Ten days later, Matheson \etal (2001)
confirmed the discovery as a type IIb supernova well past maximum light,
stating ``the spectrum 
is almost identical to one of SN~1993J obtained on day 93 after explosion.''  
Using this suggestion to constrain the explosion date for the source, the optical discovery 
occurred 83 days after the explosion
(Nakano \etal 2001) and; 
 the first radio observation was 158 days after the explosion.  
Tully (1988) determined a distance of 18.7 Mpc (assuming H$_{\rm 0}$ = 75 km s$^{-1}$ Mpc$^{-1}$)
to NGC~5033. Accepting the Tully (1988) distance and using 
H$_{\rm 0}$ = 65 km s$^{-1}$ Mpc$^{-1}$,
we will adopt the distance to the SN 
of 21.6 Mpc, which results in an absolute visible magnitude at discovery of  $< -15$, although
it is not known what the magnitude was at maximum optical brightness (Nakano \etal 2001).  
We use the Tully distance method to maintain uniformity with our distance determinations to other radio SNe
including SN 1985L (Van Dyk \etal 1998).

Stockdale \etal (2002) detected radio emission from 
SN~2001gd on 2002 February 8.54 with the Very Large Array 
(VLA)\footnote{The VLA telescope of the National Radio Astronomy Observatory is operated by Associated
Universities, Inc. under a cooperative agreement with the National Science Foundation.}
 at 22.485 GHz ($\lambda$ 1.3 cm), 8.435 GHz ($\lambda$ 3.6 cm), 
4.885 GHz ($\lambda$ 6.1 cm), and 1.465 GHz
($\lambda$ 21 cm).    The radio position of
RA(J2000) = $13^h 13^m 23\fs90$, Dec(J2000) = $+36\arcdeg 38\arcmin 18\farcs1$,
with an uncertainty of $\pm 0\farcs2$ in each coordinate (Stockdale \etal 2002), is in agreement with the optical
position to within the errors.   Since this first detection,
we have made, 
and will continue to make, radio observations with the VLA.
We note that NGC~5033 is also the host galaxy of two other historical supernovae, 
SNe~1950C and 1985L (Kowal, Sargent, \& Zwicky 1970; Aksenov \etal 1985).  
The latter was also detected in the radio in 1986 (Van Dyk \etal \ 1998).
Stockdale \etal (2002) did not detect radio
emission from either of these historical supernovae, nor do we in any of our subsequent observations.

\section{Radio Observations}

The VLA radio observations reported here were made between 2002 February 8 and 2002 October 28. 
VLA phase and flux density calibration and data reduction followed standard procedures 
(Weiler \etal \ 1986; Weiler, Panagia, \& Sramek 1990),
using 3C~286 as the primary flux density calibrator and J1310$+$323 as the phase calibrator and 
as a secondary flux density calibrator.

In Table 1, we list the flux density observations of the secondary calibrator source J1310$+$323, which are plotted in Figure 1.
The defined position for 
J1310$+$323 for phase reference is 
RA(J2000) = $13^h 10^m 28\fs 66$, Dec(J2000) = $+32\arcdeg 20\arcmin 43\farcs8$.

In Table 2, we list the flux density measurements for SN~2001gd and Figure 2 plots the data.  
The data appear to be well described by a parameterized model described below and illustrated
in Figure 2 .   

The flux density measurement error given for the VLA measurements in Table 2 is a combination of the
rms map error, which measures the contribution of small unresolved 
fluctuations in the background emission
and random map fluctuations due to receiver noise, and a basic fractional error, $\epsilon$,
included to account for the normal inaccuracy of VLA flux density
calibration and possible deviations of the primary calibrator from an
absolute flux density scale.  The final errors ($\sigma_f$) as listed
in Table 2 are taken as

\begin{equation}
\sigma_{f}^{2}\equiv(\epsilon S_0)^2+\sigma_{0}^2 
\label{eq:err}
\end{equation}

\noindent where $S_0$ is the measured flux density, $\sigma_0$ is the map rms for each
observation, and $\epsilon =$ 0.10 at  21 cm, 0.05 at 6 cm, 0.05 at 3.6 cm, 0.075 at 2 cm, and 
0.10 at 1.2~cm.

\section{Discussion}

Common properties of radio SNe (RSNe) include non-thermal
synchrotron emission with high brightness temperature, turn-on delay at longer wavelengths,
power-law decline after maximum with index $\beta$, and spectral index $\alpha$ asymptotically
decreasing to an optically thin value (Weiler \etal 1986).  Weiler \etal (1986, 1990) have shown that the ``mini-shell'' model
of Chevalier (1982a; 1982b),  adequately describes
previously known RSNe.  In this model, the relativistic electrons and enhanced magnetic fields necessary
for synchrotron emission are generated by the SN shock interacting with a relatively high-density 
envelope of circumstellar material (CSM)
which has been ionized and heated by the initial UV/X-ray flash.  This dense cocoon is presumed to have
been established by a constant mass-loss (${\dot M}$) rate, in a
constant velocity wind ($w_{\rm wind}$; i.~e., $\rho_{\rm CSM} \propto {{\dot M}\over {w_{\rm wind} \ r^2}}$)
from a red supergiant (RSG) SN progenitor or companion.  This ionized CSM is also the 
source of the initial absorption.  The
rapid rise in radio flux density results from the shock overtaking progressively more of the 
wind matter, leaving less of it
along the line-of-sight to the observer to absorb the more slowly decreasing synchrotron 
emission from the shock region.

\subsection{Parameterized Model}

Following Weiler \etal (1986; 1990; 2001; 2002), Montes, Weiler, \& Panagia (1997), 
and Chevalier (1998), we adopt the parameterized model,
\begin{equation} 
S {\rm (mJy)} = K_1 {\left({\nu} \over {\rm 5~GHz}\right)^{\alpha}}
{\left({t - t_0} \over {\rm 1~day}\right)^{\beta}} e^{-{\tau_{\rm external}}} 
{\left({1 -e^{-{\tau_{\rm CSM_{clumps}}}} \over {\tau_{\rm CSM_{clumps}}}}\right)}
{\left({1 - e^{-{\tau_{\rm internal}}}} \over {\tau_{\rm internal}}\right)}
\end{equation}
with 
\begin{equation}
\tau_{\rm external} = \tau_{\rm CSM_{homogeneous}} + \tau_{\rm distant}
\end{equation}
where
\begin{equation} 
\tau_{\rm CSM_{homogeneous}} = K_2 {\left({\nu} \over {\rm 5~GHz}\right)^{-2.1}} {\left({t -
t_0} \over {\rm 1~day}\right)^{\delta}},
\end{equation}
\begin{equation} 
\tau_{\rm CSM_{clumps}} = K_3 {\left({\nu} \over {\rm 5~GHz}\right)^{-2.1}} {\left({t -
t_0} \over {\rm 1~day}\right)^{\delta^{\prime}}},
\end{equation}
and
\begin{equation} 
\tau_{\rm distant} = K_4 {\left({\nu} \over {\rm 5~GHz}\right)^{-2.1}},
\end{equation}
\noindent with $K_1$, $K_2$, $K_3$, and $K_4$  corresponding, formally, to the flux
density ($K_1$), homogeneous absorption ($K_2$, $K_4$), and clumpy (or filamentary) absorption ($K_3$)
at 5~GHz one day after
the explosion date $t_0$.  The terms $\tau_{\rm CSM_{homogeneous}}$ and $\tau_{\rm CSM_{clumps}} $
describe the attenuation of a local homogeneous and clumpy CSM that are near enough to the supernova
progenitor that they are affected by the expanding blastwave.  The $\tau_{\rm CSM_{homogeneous}}$ 
absorption term is due to an ionized medium which uniformly covers the emitting
source. The $\tau_{\rm CSM_{clumps}}$ absorption term describes the attenuation
produced by an inhomogeneous medium (see Natta \& Panagia (1984) for a more
detailed discussion).   The $\tau_{\rm distant}$ term describes the attenuation
produced by a homogeneous medium that uniformly covers the source but is so far from the SN progenitor
that it is not affected by the blastwave and is constant  in time.

All external and clumpy absorbing media
are assumed to be purely thermal, singly ionized gas that absorbs via free-free (f-f) transitions with
frequency dependence $\nu^{-2.1}$ in the radio.  The parameter $\delta$ describes the time
dependence of the optical depth for the local, uniform medium.  
For an undecelerated SN shock, $\delta = -3$ is 
appropriate\footnote{Weiler \etal (1986) show that the absorption term 
$\tau_{\rm CSM_{homogeneous}} \propto r^{-3}$.  Since $r \propto t^1$ for an undecelerated shock,
$\tau_{\rm CSM_{homogeneous}} \propto t^{-3}$ and from Equation 4, it follows that $\delta = -3$
in this case.}, and we have assumed
this value for our parameterization since the data are insufficient to properly determine $\delta$
at this time.
The parameter $\delta^{\prime}$ describes the
time dependence of the optical depth for the non-uniform (clumpy) absorbing medium.
For this source and at this epoch, we neglect the ($K_4$) term because any 
distant, time-independent absorption effects will likely
be evident only at lower frequencies and later times than are available in our current data set
(see e.g. Montes \etal 1997; Schlegel \etal 1999 for SN~1978K).

Chevalier (1998)  proposed that  synchrotron self-absorption (SSA)
may be a significant source of absorption at high frequencies and early times for some RSNe.
Given the relatively poor
sampling of the radio emission turn-on at early times, SSA is not expected to be significant and 
our parameterization of SN~2001gd is not improved by the inclusion of SSA component.

A search for the best parameter fit to Equations 2 through 6 was carried out by minimizing the reduced
chi-squared ($\chi_{\rm red}^2$).  The best-fit parameter values and comparisons with the 
type IIb SN~1993J and the other recent RSN in NGC~5033, SN~1985L, are listed in Table 3.  The resulting
model curves are plotted in Figure 2.  The errors listed for the parameter values,
determined by the ``bootstrap'' method, are at the $1\sigma$ uncertainty level.

The  fitting procedure and examination of Figure 2 indicates that the gross properties 
of the radio emission from SN~2001gd are relatively well described by a
purely thermal absorption model such as that given by Eqs. 2 --- 6, with parameter values listed in
Table 3.  The requirement for non-zero K$_2$ and K$_3$ parameter values implies the
existence of a clumpy or filamentary absorption component embedded in a relatively homogeneous
medium (on a more detailed level, this is consistent with Case 2 discussed by Weiler \etal 2001, 2002).
The distant absorption term (K$_4$) was set to zero for reasons discussed previously.  
The parameter values are
typical of values determined for other type II SNe, although they are not tightly constrained by the relatively
sparse data set.

\subsection{Comparison to Other RSNe}

It is useful to make distance independent comparisons between SN~2001gd to SN~1985L, also in NGC~5033.
Van Dyk \etal (1998) reported 
only two 6~cm detections of SN~1985L approximately 1 year after explosion, with numerous non-detections.  
SN~1985L was identified as  type IIL by Filippenko \& 
Sargent (1985) and Kris (1985).  
Despite the relatively small amount of radio data available for the SN~1985L, there are some
valid comparisons that can be made.  SN~2001gd 
($L_{\rm 6\ cm\ peak} = 2.91\times 10^{27}$ erg s$^{-1}$ Hz$^{-1}$)
 is almost ten times more luminous at the 6~cm peak flux 
density than SN~1985L 
($L_{\rm 6\ cm\ peak} = 3.10\times 10^{26}$ erg s$^{-1}$ Hz$^{-1}$; Van Dyk \etal 1998).\footnote{We have adjusted the luminosity of SN~1985L to account for a Tully (1988)  distance
to NGC~5033 of 21.6, assuming H$_{\rm 0}$ = 65 km s$^{-1}$ Mpc$^{-1}$,  Van Dyk \etal (1998) 
assumed H$_{\rm 0}$ = 70 km s$^{-1}$ Mpc$^{-1}$ for a Tully (1988) distance of 20 Mpc.} 
The 6~cm peak luminosity of SN~2001gd also lies within a range of other type II SNe, for example, 
SN~1979C ($2.55\times 10^{27}$ erg s$^{-1}$ Hz$^{-1}$) and SN~1980K
 ($1.18\times 10^{26}$ erg s$^{-1}$ Hz$^{-1}$).
SNe~2001gd and 1985L also have very different
times from explosion to 6~cm peak, although in both cases there is great uncertainty in this time scale.  
SN~2001gd has a much steeper spectral index ($\alpha = -1.4$) than any other well-observed RSN. 
Weiler \etal (2002) demonstrate that most type II SNe have a spectral index ($\alpha > -1.0$),
similar to SN~1985L which has a moderate value of $\alpha = -0.58$.  The power law decline index for 
SN~2001gd ($\beta \sim -0.96$) is comparable to that of other type II RSNe.

It is also useful to make some comparisons between SNe~2001gd and 1993J since both are type IIb
SNe and SN~1993J was much better sampled than SN~2001gd.  Both have 
relatively equivalent 6~cm peak luminosities, as reported in Table 3.
The time to 6~cm peak for SN~1993J is in relatively close agreement with SN~2001gd ($\sim 170$
days), although this value is somewhat atypical when compared with other type II SNe.
Type Ib/Ic SNe typically achieve their 6~cm peak
luminosity in $<$40 days, while most type II SNe take from as a few as $\sim$100 days to well over 
1 year (Weiler \etal 2002).  Both SNe 2001gd and 1993J have spectral indices $< -1.0$, which is more typical
of type Ib/Ic SNe (Weiler \etal 2002).
But, SN~2001gd appears to have a power-law decline index 
comparable to that of SN~1993J and other type II SNe, unlike
values reported for type Ib SNe~1983N and 1984L ($\beta \sim -1.6$;
Sramek, Panagia, \& Weiler 1984; Weiler \etal 1986; Panagia, Sramek, \&Weiler 1986).  

%

\subsection{Estimates of Mass-loss Rates}

Using the derived parameter values from the radio light curves, 
we can estimate the presupernova mass loss rate of the SN~2001gd progenitor star 
 (Weiler, Panagia, \& Montes, 2001; Weiler \etal 1986; 2001; 2002), as

\begin{eqnarray}
{{\rm \dot M/(M_{\sun}\ {\rm yr}^{-1})}\over{w_{\rm wind}/{\rm (10\ km\ s^{-1}})}} & = & 3.0\times 10^{-6} \left<{\tau_{\rm eff}^{0.5}}\right> m^{-1.5}\left({{v_{i}}\over{\rm 10^4 \ km \ s^{-1}}}\right)^{1.5}  \nonumber \\
 &\times & \left({{t_{\rm i}}\over{\rm 45 \ days}}\right)^{1.5} \left({{t}\over{t_{\rm i}}}\right)^{1.5m} \left({{T}\over{\rm 10^4 \ K}}\right)^{0.68},
\end{eqnarray}

where

\begin{equation}
\left<{\tau_{\rm eff}^{0.5}}\right> = 0.67 [(\tau_{\rm CSM_{\rm homogeneous}} + \tau_{\rm CSM_{\rm clumps}})^{1.5}
- \tau_{\rm CSM_{\rm homogeneous}}^{1.5}]\tau_{\rm CSM_{\rm clumps}}^{-1}
\end{equation}.

This estimate depends on the parameterized model as well as on assumptions of temperature 
($T_{\rm CSM} = 2\times10^4 \ {\rm K}$), pre-SN wind speed ($w_{\rm wind} = {\rm 10 \ km \ s^{-1}}$), 
shock velocity ($v_{i} = 10^4 \ {\rm km \ s^{-1}}$), and $t_i$ which accounts for the time between
the appearance of optical lines from which the expansion velocity is determined, and the explosion date of the SN, 
typically assumed to be 45 days (Weiler \etal 1986).   Equation 7 is derived from 
Case 2\footnote{Case 1 is a uniform, external absorbing medium; Case 2 is a clumpy or filamentary
medium with a statistically large number of clumps; and Case 3 is similar to Case 2, but with a 
statistically small number of clumps.} presented by Weiler \etal (2001; 2002).
Since the SN was detected at such a late time, we can only
approximately estimate the shock velocity and temperature at  early times, 
making the mass loss rate estimates for SN~2001gd rather uncertain.
However, the estimated mass loss rate of $3\times 10^{-5} \ {\rm M}_{\sun}\ {\rm yr}^{-1}$
 is close to the values found for both SNe~1993J and 1985L.   
As one might expect since they are both type IIb SNe, there are many similarities between SNe~2001gd and 1993J, 
especially in that they require the presence of both uniform and clumpy (or filamentary) media to 
account for the radio light curve behavior while the mass loss rate for SN~2001gd appears to be in the lower range of
known mass loss rates  for  type II SNe ($1\times 10^{-5}$---$1\times 10^{-4} \ {\rm M}_{\sun}\ {\rm yr}^{-1}$),
 it clearly exceeds the values for
typical Ib/Ic SNe ($7\times 10^{-7}$---$1\times 10^{-5} \ {\rm M}_{\sun}\ {\rm yr}^{-1}$; Weiler \etal 2002).

\section{Conclusions}

We present  new observations of SN~2001gd at multiple radio wavelengths for the first year
after explosion.  Modeling implies that the evolution of the radio emission from SN~2001gd behaves in a
systematic fashion and is consistent with the
non-thermal emitting, thermal absorbing model previously used successfully for other RSNe.  
The CSM  required by the model fits is best described as a 
``filamentary'' or ``clumpy''  medium  embedded in a uniform component and appears similar to that
suggested for the
type IIb SN~1993J (Van Dyk \etal 1994).  In general, SN~2001gd is 
 very similar to  SN~1993J  suggesting that type IIb SNe may be relatively uniform in their properties.
By comparison, the type II-L SN~1985L in the same galaxy as SN~2001gd appears significantly different
in its radio properties, illustrating the general diversity of type II RSNe.

\acknowledgements

We are indebted to Dr.~Barry Clark at the VLA for scheduling our numerous observations and to observers
who contributed their telescope time.  KWW wishes to thank the Office of Naval Research (ONR) for the
6.1 funding supporting this research.  CJS is a National Research Council Postdoctoral Fellow.
We would like to thank A. Ray and P. Chandra for sharing their 21~cm GMRT observation.
Additional information and data on radio supernovae can be found
on {\it http://rsd-www.nrl.navy.mil/7214/weiler/} and linked pages.  We also would like to thank our
anonymous referee for his/her useful comments.

\bibliographystyle{apj}

\bibliography{References}
\clearpage
\begin{deluxetable}{cccccccc}
\tabletypesize{\footnotesize}
\tablewidth{6.75in}
\table columns(8}
\tablecaption{Calibrator Measurements for J1310$+$323}
\tablehead{
\colhead{Obs.} & \colhead{Ref. Epoch} & \colhead{Tel.} & 
\colhead{S(21cm)} & \colhead{S(6cm)} & \colhead{S(3.6cm)} & \colhead{S(2cm)} &  \colhead{S(1.2cm)} \\
\colhead{Date} &  \colhead{(days)} & \colhead{Config.} & \colhead{(Jy)} & \colhead {(Jy)}  & \colhead{(Jy)} & \colhead{(Jy)} & \colhead{(Jy)}
} 
\startdata
2001-Nov-24 & \llap{$\equiv$}  0.00  & & & & & & \\
2002-Feb-08 & 75.72 & VLA-A &  1.703 &  1.659 & 1.765   &          & 2.145   \\
2002-Mar-02 & 97.60 & VLA-A &  1.703 \tablenotemark{a}  & 1.687 & 1.745 &  2.027 & 2.164 \\
2002-Mar-14 & 109.66 & VLA-A &  1.700   &  &  &  &   \\
2002-Mar-23 & 118.42 & VLA-A &  1.681   &             &  1.798 \tablenotemark{a} & 2.077 & 2.315  \\
2002-Apr-06 & 132.38 & VLA-A &  1.690  & 1.700\tablenotemark{a}  & 1.833\tablenotemark{a}  &  2.073 &  \\
2002-Jun-12 & 199.30 & VLA-B &  1.668   & 1.726 & 2.005  &  &   \\
2002-Jul-16 & 234.17 & VLA-B &  1.722  & 1.773 & 2.045   &    &    \\
2002-Oct-28 & 337.15 & VLA-C &  &   & 2.130  & 2.551  & 2.728 \\
\enddata
\tablenotetext{a}{Interpolated from other calibrator measurements.}
\end{deluxetable}

\clearpage
\begin{deluxetable}{cccccccc}
\tabletypesize{\scriptsize}
\tablewidth{6.75in}
\table columns(8}
\tablecaption{Flux Density Measurements for SN~2001gd}
\tablehead{
\colhead{Obs.} & \colhead{Ref. Epoch} & \colhead{Tel.} & 
\colhead{S(21cm) $\pm \ \sigma_f$} & \colhead{S(6cm) $\pm \ \sigma_f$} & \colhead{S(3.6cm) $\pm \ \sigma_f$} & \colhead{S(2cm) $\pm \ \sigma_f$} &  \colhead{S(1.2cm) $\pm \ \sigma_f$} \\
\colhead{Date} &  \colhead{(days)} & \colhead{Config.} & \colhead{(mJy)} & \colhead {(mJy)}  & \colhead{(mJy)} & \colhead{(mJy)} & \colhead{(mJy)}
} 
\startdata
2001-Nov-24\tablenotemark{a}  & \llap{$\equiv$}  0.00  & & & & & & \\
2002-Feb-08 & 75.72 & VLA-A &  0.848 $\pm$ 0.129 &  5.050 $\pm$ 0.257 & 4.090 $\pm$ 0.208   &                              & 1.370 $\pm$ 0.176  \\
2002-Mar-02 & 97.60 & VLA-A & 1.414 $\pm$ 0.177  &  5.414 $\pm$ 0.284 & 3.732 $\pm$ 0.195  &   2.183  $\pm$ 0.242 & 1.335 $\pm$ 0.182   \\
2002-Mar-14 & 109.66 & VLA-A & 1.840 $\pm$ 0.352 &                             &                              &                               &                            \\
2002-Mar-23 & 118.42 & VLA-A & 1.830 $\pm$ 0.339 &                             & 3.660 $\pm$ 0.208 &    2.330  $\pm$ 0.322 &  1.000 $\pm$ 0.187   \\
2002-Apr-06 & 132.38 & VLA-A & 1.990 $\pm$ 0.344  &  5.330 $\pm$ 0.277 & 3.780 $\pm$ 0.212 &   2.660  $\pm$ 0.304\tablenotemark{b} &              \\
2002-Jun-12 & 199.30 & VLA-B & 3.230 $\pm$ 0.365  & 4.842 $\pm$ 0.249  &  3.100 $\pm$ 0.168 &                              &                    \\
2002-Jul-16 & 234.17 & VLA-B & 2.733 $\pm$ 0.314  &  4.260 $\pm$ 0.234   & 2.390 $\pm$ 0.137 &                               &                       \\
2002-Sep-22 & 301.40 & GMRT & 3.550 $\pm$ 0.220\tablenotemark{c}   &                                &                            &                              &                    \\
2002-Oct-28 & 337.15 & VLA-C &                            &                                & 2.071 $\pm$ 0.136 & 0.893 $\pm$ 0.178   &  0.809 $\pm$ 0.155  \\
\enddata
\tablenotetext{a}{Nakano \etal (2001); Matheson \etal (2001); SN~2001gd is believed to have exploded roughly 83 days 
prior to its initial optical detection.  See Section 1 for further explanation.}
\tablenotetext{b}{This value was not included in the modeling.}
\tablenotetext{c}{Chandra, Ray, \& Bhatnagar (2002)}

\end{deluxetable}

\clearpage

\begin{deluxetable}{cccc}
\tabletypesize{\footnotesize}
\tablecaption{Fit  of SN~2001gd \& Comparison with other Type II Supernovae. \label{tbl-3}}
\tablewidth{0pt}
\tablehead{
\colhead{Parameter\tablenotemark{a}} & \colhead{SN~2001gd} & \colhead{SN~1993J\tablenotemark{b}}
& \colhead{SN~1985L\tablenotemark{c}}}
\startdata
$K_1$                 & $\left({1.49^{+1.88}_{-1.09}}\right) \times 10^3$       & $4.14\times 10^3$ & $4.8 \times 10^3$  \\
$\alpha$             & $-1.38^{+0.13}_{-0.12}$                                       & $-0.99$              & $>-0.58$                   \\
$\beta$              & $-0.96^{+0.24}_{-0.13}$                                      & $-0.64$                & $<-0.71$                   \\ 
$K_2$                 & $\left({3.25^{+4.55}_{-1.31}}\right) \times 10^6$      & $1.35 \times 10^3$ &  $>7.00 \times 10^6$ \\
$\delta$              & $\equiv -3.00$                                                  &   $-1.99$               &   $\equiv -3.00$       \\
$K_3$                & $\left({1.05^{+1.31}_{-0.95}}\right) \times 10^3$   & $2.54 \times 10^4$     & $0.00$                    \\
$\delta^{\prime}$  & $-1.27^{+0.40}_{-0.15}$                                  & $-2.02$                   & $0.00$                            \\ 
Time to $L_{\rm 6\ cm\ peak}$ \ (days) & $173^{+10}_{-15}$            & 168                        & $\equiv 30$  \\
$L_{\rm 6\ cm\ peak}$\ (erg s$^{-1}$ Hz$^{-1}$) & $\left({2.91^{+0.13}_{-0.04} }\right) \times 10^{27}$  &  $1.45 \times 10^{27}$ & $\sim 3.1 \times 10^{26}$         \\
${\rm \dot M}$ (${\rm M_\odot}$ yr$^{-1}$) & $3.0 \times 10^{-5}$ & $2.1 \times 10^{-5}$ & $>6.3 \times 10^{-5}$  \\
\enddata
\tablenotetext{a}{Assumes a constant mass-loss rate, constant velocity, wind-established CSM, i.e., $\rho \propto r^{-2}$.}
\tablenotetext{b}{Van Dyk \etal (1994)}
\tablenotetext{c}{Van Dyk \etal (1998)}
\end{deluxetable}

\clearpage

\figcaption[sn01gd-calplot2.eps]{Plot of the flux denstity of the VLA calibrator source J1310$+$323 at 1.3 cm ({\it circles}), 2 cm ({\it crosses}), 3.6 cm ({\it squares}), 
6 cm ({\it triangles}), and 21 cm ({\it diamonds}).  Note that the values for 21 cm are all reduced by 0.5 Jy to avoid confusion on the plot.  \label{fig1}}

\figcaption[rsnplot01gd_apj.ps]{The radio measurements and parameterized model for SN~2001gd at 1.3 cm ({\it circles; solid line}), 2 cm ({\it crosses; dashed line}), 3.6 cm ({\it squares; dash-dot line}), 
6 cm ({\it triangles; dotted line}), and 21 cm ({\it diamonds; dash-triple dotted line}).  The lines represent the best fit model as described in the text with the parameters listed in Table 2.
SN~2001gd is believed to have exploded 83 days prior to its initial optical discovery, see Section 1 for
further explanation(Nakano \etal 2001; Matheson \etal 2001). \label{fig2}}

\clearpage

\begin{figure}
\figurenum{fig1}
\epsscale{1.0}
\plotone{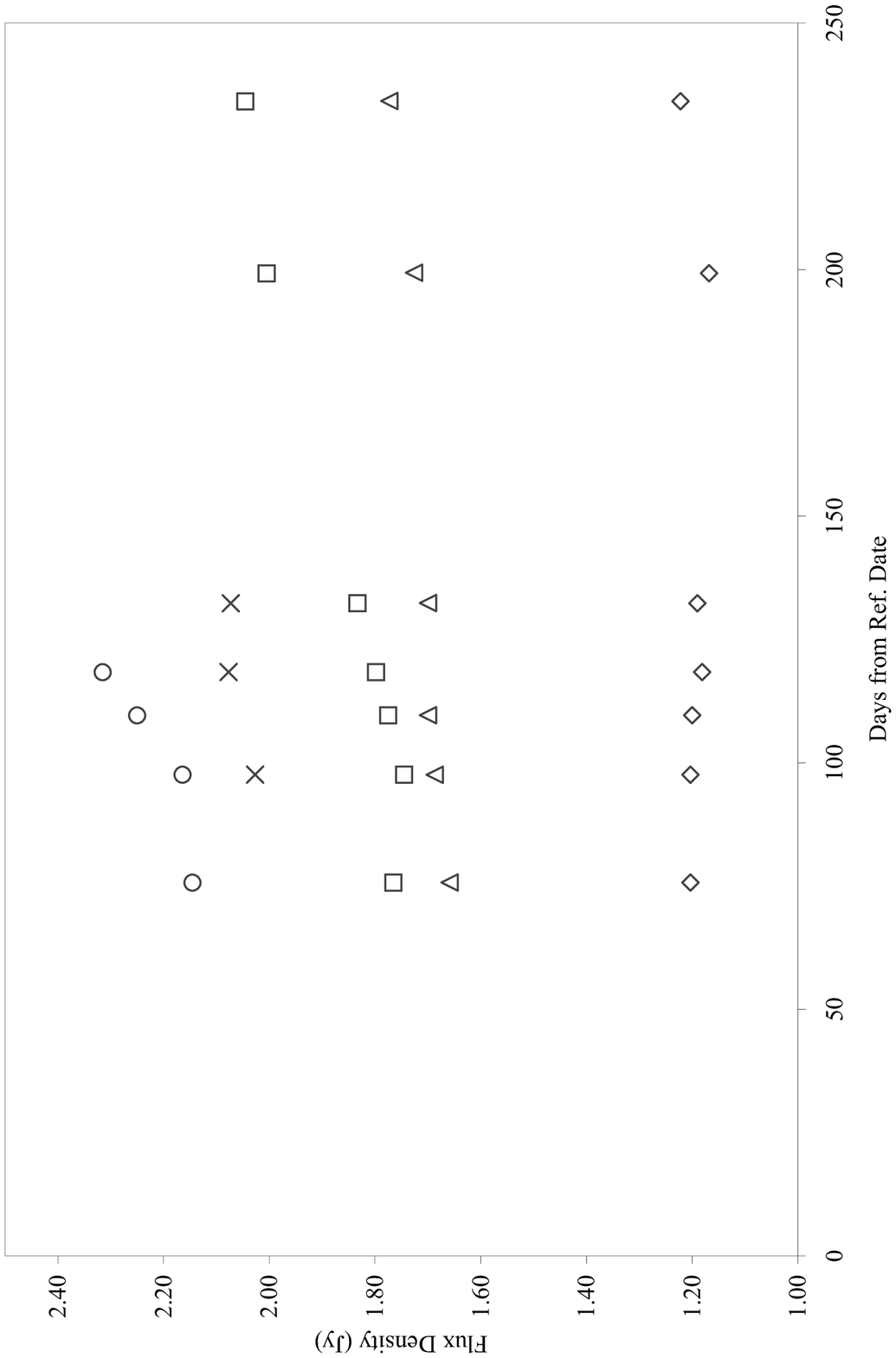}
\end{figure}
 
\clearpage

\begin{figure}
\figurenum{fig2}
\epsscale{1.0}
\plotone{f2.ps}
\end{figure} 

\clearpage

\end{document}